\documentclass[conference]{IEEEtran}
\IEEEoverridecommandlockouts
\usepackage{cite}
\usepackage{amsmath,amssymb,amsfonts}
\usepackage{algorithmic}
\usepackage{graphicx}
\usepackage{textcomp}
\usepackage{xcolor}
\usepackage{subcaption}
\usepackage{hyperref}
\usepackage{booktabs}
\usepackage{float}
\usepackage{multirow}
\restylefloat{table}
\def\BibTeX{{\rm B\kern-.05em{\sc i\kern-.025em b}\kern-.08em
    T\kern-.1667em\lower.7ex\hbox{E}\kern-.125emX}}
\begin{document}

\title{It's All About The Cards: Sharing on Social Media Probably Encouraged HTML Metadata Growth}

\author{
    \IEEEauthorblockN{
        Shawn M. Jones\IEEEauthorrefmark{1}\IEEEauthorrefmark{3},
        Valentina Neblitt-Jones\IEEEauthorrefmark{1}\IEEEauthorrefmark{4},  
        Michele C. Weigle\IEEEauthorrefmark{2}\IEEEauthorrefmark{5},
        Martin Klein\IEEEauthorrefmark{1}\IEEEauthorrefmark{6}, and
        Michael L. Nelson\IEEEauthorrefmark{2}\IEEEauthorrefmark{7}
        }
    \IEEEauthorblockA{
        \IEEEauthorrefmark{1}Los Alamos National Laboratory, Los Alamos, New Mexico
    }
    \IEEEauthorblockA{
        \IEEEauthorrefmark{2}Old Dominion University, Norfolk, Virginia
    }
    \IEEEauthorblockA{
        \IEEEauthorrefmark{3}smjones@lanl.gov, 0000-0002-4372-870X
    }
    \IEEEauthorblockA{
        \IEEEauthorrefmark{4}vneblitt@lanl.gov, 0000-0002-6881-4268
    }
    \IEEEauthorblockA{
        \IEEEauthorrefmark{5}mweigle@cs.odu.edu, 0000-0002-2787-7166
    }
    \IEEEauthorblockA{
        \IEEEauthorrefmark{6}mklein@lanl.gov, 0000-0003-0130-2097
    }
    \IEEEauthorblockA{
        \IEEEauthorrefmark{7}mln@cs.odu.edu, 0000-0003-3749-8116
    }

}

\maketitle

\begin{abstract}
In a perfect world, all articles consistently contain sufficient metadata to describe the resource. We know this is not the reality, so we are motivated to investigate the evolution of the metadata that is present when authors and publishers supply their own. Because applying metadata takes time, we recognize that each news article author has a limited metadata budget with which to spend their time and effort. How are they spending this budget? What are the top metadata categories in use? How did they grow over time? What purpose do they serve? We also recognize that not all metadata fields are used equally. What is the growth of individual fields over time? Which fields experienced the fastest adoption? In this paper, we review 227,726 HTML news articles from 29 outlets captured by the Internet Archive between 1998 and 2016. Upon reviewing the metadata fields in each article, we discovered that 2010 began a metadata renaissance as publishers embraced metadata for improved search engine ranking, search engine tracking, social media tracking, and social media sharing. When analyzing individual fields, we find that one application of metadata stands out above all others: social cards --- the cards generated by platforms like Twitter when one shares a URL. Once a metadata standard was established for cards in 2010, its fields were adopted by 20\% of articles in the first year and reached more than 95\% adoption by 2016. This rate of adoption surpasses efforts like Schema.org and Dublin Core by a fair margin. When confronted with these results on how news publishers spend their metadata budget, we must conclude that it is all about the cards.
\end{abstract}


\section{Introduction}

Metadata is key to organizing content and providing important context. Schriml et al. \cite{schriml_covid-19_2020} recently highlighted how the lack of metadata may have impacted the ability of researchers to respond to the COVID-19 pandemic. What about the historians of the future? When they review the news stories and other information about the COVID-19 pandemic, what metadata will they have to provide them context?

Creating content requires time and effort. Metadata is often created once the content is complete. In traditional library settings metadata is generated by the publisher of the work as well as librarians and archivists, sometimes years later. The drive to quickly release content on the web, especially with news stories, leads publishers to be judicious about the amount of time and effort they expend creating metadata for their publications. Thus, web publication processes have a metdata budget. Once the participants in the process exhaust that budget, likely because a deadline is looming, they can expend no more effort on metadata. The metadata with the highest impact becomes the focus of the effort. So, how have different news publishers spent their metadata budget?


In this paper, we show how metadata that supports specific functions has encouraged a metadata renaissance. We sampled from the NEWSROOM dataset \cite{grusky_newsroom_2018} to acquire 277,724 HTML news articles from 29 outlets captured by the Internet Archive between 1998 through 2016. We analyzed the rise of different metadata standards used with the HTML \texttt{META} element. We find that the mean number of metadata fields in use in 1998 is two. Throughout the dataset, roughly two metadata fields are added per year, reaching a mean of 39 metadata fields per page in 2016. Since 2008, we see an explosion of metadata usage for the purposes of advertising, browser customization, search engine verification, and social media sharing. The metadata categories with the highest growth are those for social media.

\begin{figure}[t]
    \centering
    \includegraphics[width=0.4\textwidth]{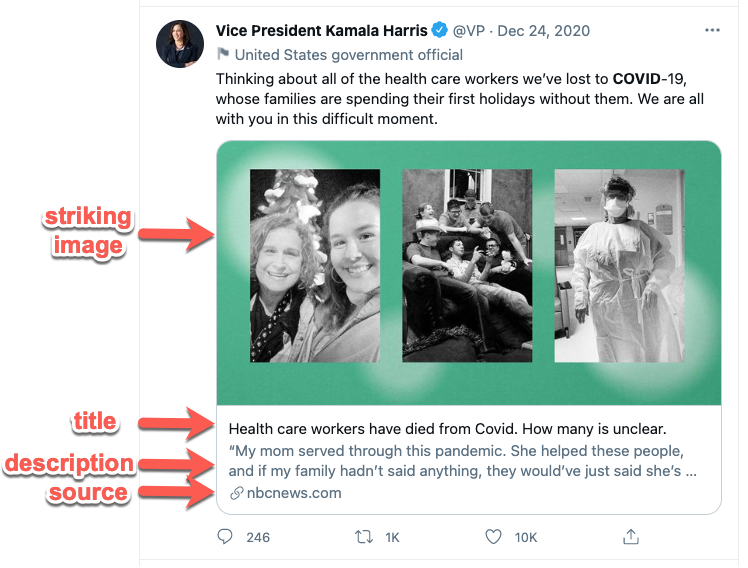}
    \caption{Social cards are routinely shared on Twitter and Facebook. They contain a striking image, a title, a description, and source attribution.}
    \label{fig:twitter_annotated_social_card}
\end{figure}

We further analyzed the top categories to discover the usage and growth of individual fields. We found the highest growth with fields dedicated not just to social media, but specifically to \textbf{social cards}, surrogates that describe individual resources on social media. An example of a Twitter social card is shown in Figure \ref{fig:twitter_annotated_social_card}. These surrogates are similar to search engine snippets, but serve a slightly different purpose. Where search engine snippets are generated dynamically based on a user's query and try to answer the question \emph{Will this resource meet my information need?}, social cards try to answer the question \emph{What does the resource contain?} Social cards typically consist of a striking image, title, description, and source attribution. In prior work \cite{jones_social_2019}, we demonstrated that groups of social cards perform best for understanding collections. Social cards have become standard currency in the sharing culture, existing not only on social media platforms, but also in messaging apps like Apple Messages, storytelling services like Wakelet\footnote{\url{https://wakelet.com/}}, and aggregation platforms like Flipboard\footnote{\url{https://flipboard.com/}}.


News articles represent resources that have undergone editorial review, receiving some care in their publication. Thus, with these articles, we answer the following research questions:

\textbf{Research Question \#1 (RQ1)} --- What are the top metadata categories, their purpose, and how do their frequencies of use evolve over time?

\textbf{Research Question \#2 (RQ2)} --- Not all metadata fields are used equally; how has the adoption of specific metadata fields evolved over time?



\section{Background}
\label{sec:background}

Many works describe metadata as \emph{data about data} or \emph{information about information}. Greenberg \cite{greenberg_metadata_2003} notes that ``these definitional phrases are ambiguous given the many different uses of the terms data and information.'' Instead, she defines metadata as ``structured data about an object that supports functions associated with the designated object.'' Considering the varied functions of metadata on the web, we will use her definition in this article.

Since version 2.0 \cite{html2}, HTML has included the \texttt{META} element for suppying metadata for a web page. Each \texttt{META} element provides a key-value pair. A web page author supplies the key via the \emph{name} or \emph{property} attribute and supplies its corresponding value via the \emph{content}, \emph{value}, or sometimes \emph{href} attribute. For example, if this paper were a web page, its \emph{author} field would be displayed like the following:

{\footnotesize \texttt{<META name="author" content="Shawn M. Jones">}}

In this paper, each metadata \textbf{property} corresponds to a given concept, like \emph{title}. Each property has a corresponding \textbf{field} in a given metadata standard. For example, the \emph{title} field in Dublin Core, the \emph{name} field in Schema.org, the \emph{og:title} field in Open Graph Protocol (OGP), and \emph{twitter:title} in the Twitter Cards standard all correspond to the property of \emph{title}.

These metadata fields provided a variety of functions. Some, like \emph{author}, provide descriptive metadata for the article. Others, like \emph{robots}, exist to instruct search engines on how to crawl the page. Still others, like \emph{fb:app\_id} and \emph{msvalidate.01}, provide an identifier so that web platforms can provide analytics and insight. Because we are interested in the functions of widely used metadata, we do not confine ourselves to just descriptive metadata fields.

HTML provides a list of standard metadata names \cite{html_meta} for certain document properties. They are \emph{application-name}, \emph{author}, \emph{description}, \emph{generator}, and \emph{keywords}. The WhatWG standard \cite{whatwg_meta} also mentions \emph{referrer}, \emph{theme-color}, and \emph{color-scheme}. Some, like \emph{application-name} do not apply to all types of documents. Others, like \emph{theme-color} and \emph{color-scheme} are not descriptive metadata, but instead specifically used to inform the browser about how the page should be rendered. To differentiate them from other metadata standards or use cases, we refer to these fields collectively as \textbf{Standard HTML Metadata}.

The Dublin Core Element Set \cite{dc_element_set} (DC) was originally conceived in 1995 to create a minimal set of properties for resource description. It was designed to be generic enough to apply to many different types of resources, from web pages to videos to physical objects. It consists of 15 properties: \emph{contributor}, \emph{coverage}, \emph{creator}, \emph{date}, \emph{description}, \emph{format}, \emph{identifier}, \emph{language}, \emph{publisher}, \emph{relation}, \emph{rights}, \emph{source}, \emph{subject}, \emph{title}, and \emph{type}. These original 15 properties have been extended in the new DCMI Metadata Terms standard \cite{dc_metadata_terms} to allow page authors to support more granular concepts, such as \emph{dateSubmitted} and \emph{dateCopyrighted} rather than just \emph{date}. DCs properties are established in the \texttt{META} element through the prefix \emph{dc.} or \emph{dcterms.}

\begin{table}[t]
    \caption{Social card units and their associated cards standards fields}
    \label{tab:social_card_units}
    \scriptsize
    \begin{tabular}{@{}lll@{}}
    \toprule
    \textbf{Social Card Unit}    & \textbf{OGP}   & \textbf{Twitter Card} \\ \midrule
    title                        & \emph{og:title}       & \emph{twitter:title}         \\
    description                  & \emph{og:description} & \emph{twitter:description}   \\
    striking image               & \emph{og:image}       & \emph{twitter:image}         \\
    identify the resource        & \emph{og:url}         & N/A                   \\
    specify the type of resource & \emph{og:type}        & \emph{twitter:card}          \\ \bottomrule
    \end{tabular}%
\end{table}

\begin{table*}[t]
    \caption{Studies performed to understand Dublin Core usage for different types of documents.}
    \label{tab:dublin_core_studies}
    \centering
    \scriptsize
    \begin{tabular}{@{}lllll@{}}
    \toprule
    \textbf{Study}              & \textbf{Publication} & \textbf{Document} & \textbf{Top-5 DC} & \textbf{Top-5 non-DC} \\                       
    & \textbf{Year} & \textbf{Type} & \textbf{Fields Discovered} & \textbf{Fields Discovered}                    \\ \midrule
    O'Neill \cite{oneill_web_2001}            & 2001             & web pages                           & title, description, subject, language, publisher    & generator, keywords, content-type, description, author    \\ \midrule
    Ward \cite{ward_quantitative_2003}              & 2003             & OAI-PMH sites                       & creator, identifier, title, date, type              &                                                           \\ \midrule
    Park \cite{park_semantic_2006}              & 2006             & image collections                   & subject, description, title, format, coverage       &                                                           \\ \midrule
    Alijani and Jowkar \cite{saadat_alijani_dublin_2009} & 2009             & web pages                           & title, publisher, language, creator, date           &                                                           \\ \midrule
    Park and Richard \cite{park_metadata_2011}  & 2010             & electronic theses & title, subject, desciption, rights, type            &                                                           \\ 
    & & and dissertations \\ \midrule
    Weagley \cite{weagley_interoperability_2010}           & 2010             & video repositories                  & title, description, date, identifier, type          &                                                           \\ \midrule
    Ard{\"o} \cite{ardo_can_2010}               & 2010             & web pages                           & title, language                                     & keywords, description, author, content language, language \\ \midrule
    Phelps \cite{phelps_evaluation_2012}            & 2012             & web pages                           & date, title, language, creator, subject             & content-type, keywords, description, robots, generator    \\ \midrule
    Bu and Park \cite{bui_assessment_2013}       & 2013             & OAI-PMH sites                       & title, identifier, date, description, creator       &                                                           \\ \midrule
    Windnagel  \cite{windnagel_usage_2014}        & 2014             & math and science       & description, identifier, contributor, title, format &                                                           \\ 
    & & repositories & & \\
    \bottomrule
    \end{tabular}
\end{table*}

\begin{table}[b]
    \caption{NEWSROOM sample data reduction for metadata availability analysis}
    \centering
    \label{tab:newsroom_sample_analysis}
        \footnotesize
    \begin{tabular}{@{}lrr@{}}
    \toprule
                            & \textbf{Count}  & \textbf{Running Total} \\ \midrule
    Initial Sample          & 310,163 & 310,163        \\
    Connection Failures     & 734    & 309,429        \\
    404 Not Found           & 7,570   & 301,859        \\
    503 Service Unavailable & 355    & 301,504        \\
    429 Too Many Requests   & 110    & 301,394        \\
    405 Method Not Allowed  & 54     & 301,340        \\
    403 Forbidden           & 2      & 301,338        \\
    400 Bad Request         & 1      & 301,337        \\
    Processing failures & 4,447 & 296,890 \\
    Redirects to dates after 2016 & 19,164 & 277,724 \\
    Remaining For Analysis  &        & 277,724        \\ \bottomrule
    \end{tabular}
\end{table}


In 2010, Facebook established the Open Graph Protocol (OGP) \cite{facebook_open_2021} standard for describing web resources shared through social media. Twitter followed with its Twitter Card \cite{twitter_cards_2021} standard. With these standards, web page authors apply the appropriate field name to an HTML \texttt{META} element on their page to fill in the appropriate card unit, as shown in Table \ref{tab:social_card_units}. Figure \ref{fig:twitter_annotated_social_card} displays a card for a CNN news article where all of the appropriate fields were specified. Whereas the Twitter Card standard focuses primarily on cards, OGP also supports 60 metadata fields for card generation, user tracking, and advertising purposes. When generating a social card, if the Twitter Card fields are not present, Twitter will fallback to using the OGP equivalents, but the page must still contain the entry \emph{twitter:card}.

The Schema.org standard \cite{guha2016schema} was developed in 2011 as a joint venture between Google, Microsoft, Yahoo, and Yandex. The goal of this standard is to promote the description of items on a web page so that search engines can build better models of aboutness. This allows search engines to group pages selling the same product, or pages by the same author or publisher, and provide insights into the content, such as prices or topics. Hendler \cite{hendler_peta_2013} noted in 2013 that Schema.org adoption was driven by the perception that pages that failed to adopt it would be ranked lower in search results. He also mentioned that Schema.org is a good start toward a standard for semantic web interoperability, but does not contain the nuances necessary for a more fine grained understanding of content or relationships between facts discovered in web pages.

We retrieved our news articles from the Internet Archive. When discussing web archives, we use the terms provided by the Memento Protocol \cite{van_de_sompel_rfc_2013}. Web archives capture current web resources, or \textbf{original resources}, identified by a \textbf{URI-R}. Each capture, or \textbf{memento}, identified by a \textbf{URI-M}, is an recording of that web resource from a specific moment in time, its \textbf{memento-datetime}. Such captures are important because content changes on the web \cite{klein_scholarly_2014,jones_scholarly_2016}. Mementos represent observations of resources at key points in time, capturing the article's HTML, allowing us to evaluate the behavior of authors and publication platforms over time. The memento-datetime is not the publication date, as a web page is often captured after its content is published. For example, consider a scenario where a news article was published in 2004, but was not captured until 2014. The HTML captured in 2014 represents the behavior of a publication platform from 2014, \emph{not 2004}. In this paper we are most interested in web page publisher behavior, so we rely upon memento-datetime rather than publication date to understand how publications applied metadata over time. Because our study examines different metadata fields, we have to consider not only the standards as they currently exist, but all versions prior. This can make for a very confusing comparison when fields have changed over time. For example, \emph{og:latitude} was used by some articles between 2010 and 2016. Is it user error or the application of an archaic field?

We used the NEWSROOM \cite{grusky_newsroom_2018} dataset developed by Grusky et al. The NEWSROOM dataset consists of JSON Lines (JSONL) files containing 1.3 million records of news articles from 29 news outlets captured between 1998 and 2016 by the Internet Archive. The dataset's original intention was to provide input for evaluating automatic text summarization algorithms. Each record consists of the article's URI-M, its title, its extracted text, a summary written by the article's author or editor extracted from its \texttt{META} element, and some derived metrics describing the nature of the summary. As a source of summaries, each news article's memento is guaranteed to contain at least a \emph{description}, \emph{twitter:description}, or \emph{og:description} field.

\section{Related Work}




In 1998, Marchiori \cite{marchiori_limits_1998} noted that metadata adoption would be key to helping search engines process and understand the web, and introduced a method to improve the generation of metadata for pages. That same year Brin and Page \cite{brin_anatomy_1998} published work on a search engine prototype named Google. Brin and Page stated that ``metadata efforts have largely failed with web search engines'' due to page authors abusing metadata to improve their search results. In 2002 Monika Henzinger had stated in an interview \cite{henzinger_perspectives_2002} that Google's operations did not universally trust page metadata because they did not want Google's results to be manipulated. She did note, however, that for known trusted sites, they might incorporate page metadata into their process. Mohamed \cite{mohamed_impact_2006} was able to leverage page metadata to improve result rankings in Alta Vista, Hotbot, and Go in 2006. Papadakos et al. \cite{papadakos_exploiting_2012} developed a proof-of-concept for a new form of web search that applied static and dynamically generated metadata to improve user satisfaction with search results. We will show that other use cases besides satisfying search engines have arisen to support metadata functions associated with news articles over time.

DC was seen as a potential unifier across many types of resources, including web pages. DC was a required metadata standard for web sites running Open Archives Initiative Protocol for Metadata Harvesting\cite{oaipmh2015} (OAI-PMH). In 2003, Ward \cite{ward_quantitative_2003} quantified DC usage across 100 OAI-PMH sites. She analyzed 910,919 records from these sites and found an average of 8 DC elements per record. She found that \emph{creator}, \emph{identifier}, \emph{title}, \emph{date}, and \emph{type} were the most used fields. Table \ref{tab:dublin_core_studies} demonstrates the different studies performed to understand the usage of DC in different datasets over time. Of particular interest is Phelps' 2012 study because he compared results from Ward's 2003 study \cite{ward_quantitative_2003} and Alijani's 2009 study \cite{saadat_alijani_dublin_2009}, with his own results gathered in 2011 to note how metadata usage may have changed over time.

As noted in Section \ref{sec:background}, there are other metadata standards besides DC. Hartig \cite{hartig_provenance_2009} analyzed 1,073,218 RDF documents in 2009 and discovered DC in use by 121 documents. In contast, vocabularies FOAF and SIOC were applied to 989,263 and 127,974 documents, respectively. Mika and Potter \cite{mika_metadata_2012} processed 3,230,928,609 web pages to discover the different metadata standards in use. They found that 25.08\% applied RDFa and 22.45\% of pages applied OGP. Neither of these studies broke down the usage of individual field names. The 2020 Web Almanac \cite{http_archive_2020_2020} analyzes metadata in terms of Search Engine Optimization (SEO), focusing on the use of \emph{robots} and \emph{canonical} fields, but does not analyze all metadata fields in use. In contrast, W3Techs \cite{w3techs_usage_2021} does break down the use of OGP across the web, but only through the last 12 months, showing a 10\% growth rate for both OGP and Twitter cards between 2020 and 2021. DC has been holding steady at 1\%.

A few studies have tried to understand the reasons why some DC fields enjoy high adoption while others do not. Park and Childress \cite{park_dublin_2009} found that many metadata experts found it difficult to apply DC fields consistently and accurately. The experts stated that DC contained too many conceptual ambiguities. Also, too many elements could conceptually overlap depending on an institution's policies on semantic interoperability. Because we cannot interview the news article authors of the past, we are going to infer their motivations through their adoption of metadata standards over time.

Our work extends these studies by including more standards than DC, and, more importantly, we include the dimension of time. Where Phelps \cite{phelps_evaluation_2012} compared his results with prior work, we use web archives to actually analyze 18 years of web pages as they existed at the time of capture, providing a more consistent set of conditions for data gathering and interpretation. We do not merely analyze DC usage or OGP usage, but provide field-by-field usage for several metadata standards and use cases. If we take the top items from each DC study, we find that \emph{title} is the most popular DC field across ten studies and \emph{description} is the most popular across six studies. When considering non-DC fields, the HTML standard \emph{description} shows up in the top five of all studies that included non-DC fields. We observe that \emph{title} and \emph{description} are units of social cards.








\begin{figure}
    \centering
    \includegraphics[width=0.3\textwidth]{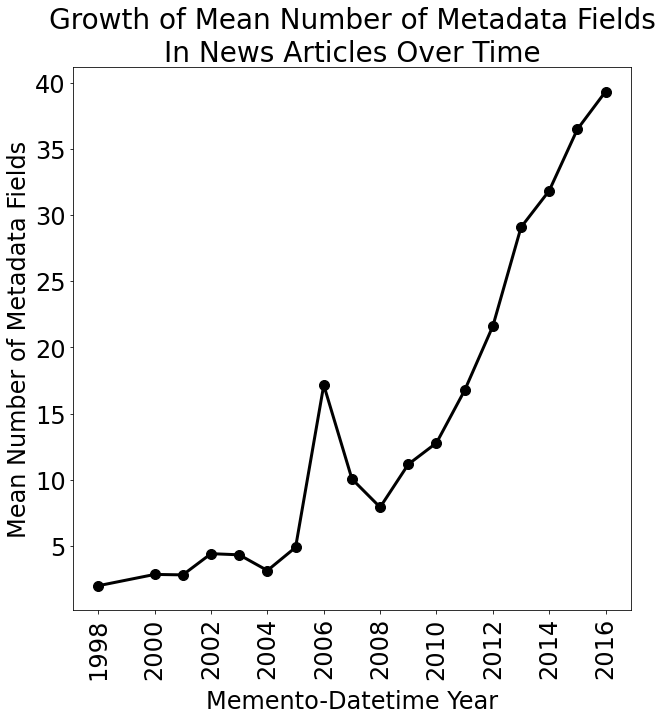}
    \caption{Growth in the mean number of metadata fields in news articles over time.}
    \label{fig:mean-metadata-fields-over-time}
\end{figure}

\begin{figure*}[t]
    \centering
    \includegraphics[width=0.7\textwidth]{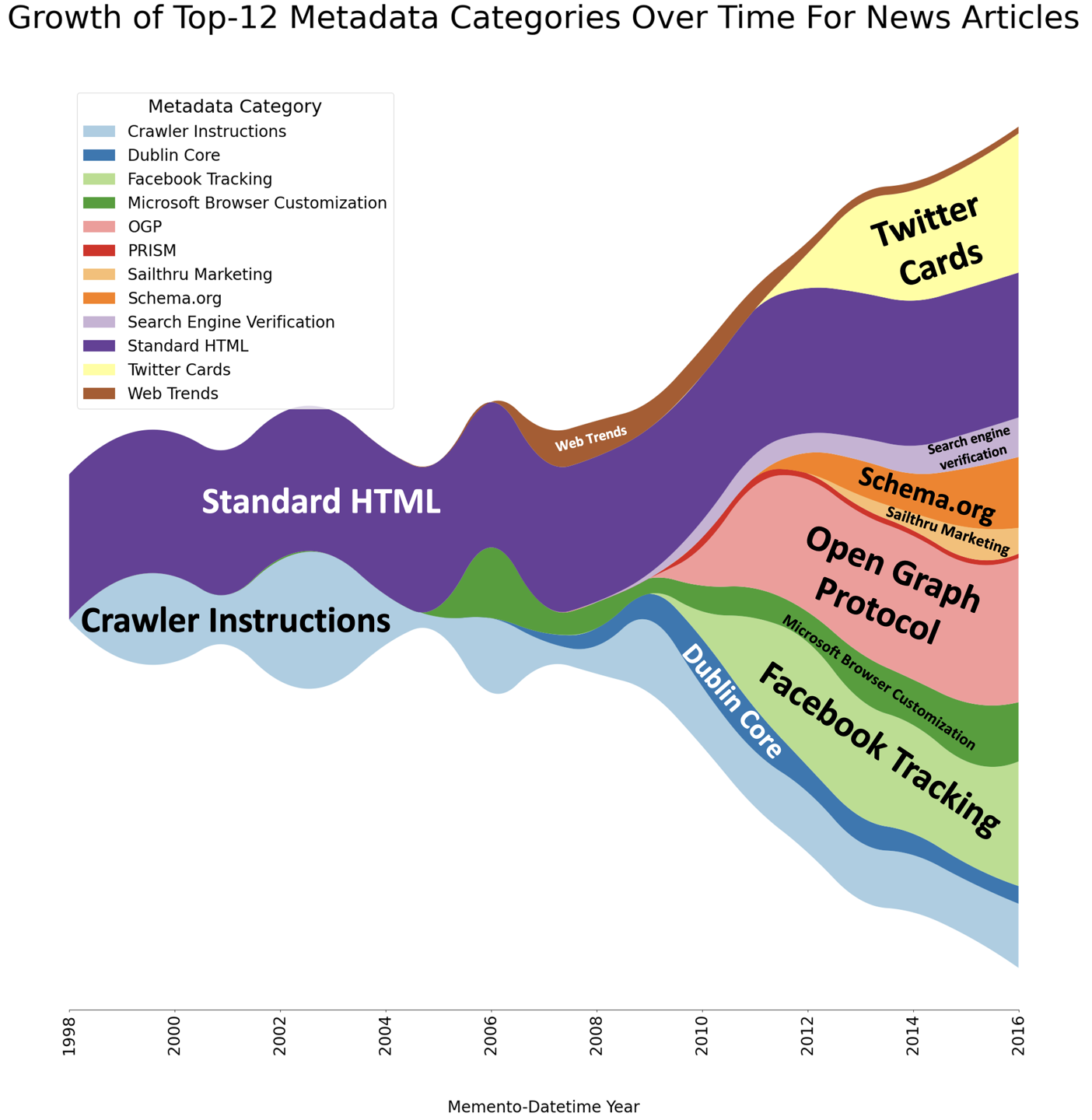}
    \caption{The top-12 metadata categories in the dataset, rendered as a streamgraph over time. The size of each category section reflects the percentage of articles from the given year that contain at least one metadata field from the given category. We see an explosion of metadata standards adoption for social media and search engines after 2008.}
    \label{fig:all_metadata_categories_streamgraph}
\end{figure*}

\section{Methodology}

We acquired the NEWSROOM dataset from Max Grusky. Even though it contains 1.3 million URI-Ms, they are not balanced with respect to domain name or memento-datetime. The dataset heavily favors \emph{nytimes.com} with 186,095 mementos coming from that domain, compared to 1,189 mementos for \emph{nbc.com}. With respect to memento-datetime year, NEWSROOM only contains 19 mementos from 1998 and 279,232 from 2016. Some 23\% of the dataset contains mementos from 2016 and the percentage decreases each year. We created a sample from the dataset that was better balanced with respect to these features. OGP was created as a social media metadata standard in 2010. To address the bias toward years closer to 2016 and to allow us to contrast metadata usage before and after social media fields were available, we included all 90,570 mementos from 2009 and before in our sample. For mementos after 2010, we randomly assigned each memento to a bucket representing its domain and year (e.g., \emph{latimes.com} in 2013). We stopped assigning mementos to a bucket once the bucket size reached 1,307 -- the median size of all domain/year divisions after 2010. This process created a more balanced sample of 310,163 mementos.

In June 2020, while downloading these mementos, we ran into a variety of issues. We divided the URI-Ms from the sample into seven subsets and downloaded each subset from servers in Amsterdam, Frankfurth, London, New York, Northern Virginia, San Francisco, and Toronto. In addition to providing parallel downloads, splitting the dataset also helped us mitigate rate limiting from the Internet Archive. The Internet Archive does not alter the appearance of mementos for different locations or languages. We repeated the downloads in July and August to account for mementos that failed to download. With the exception of mementos with a status code of 429, all mementos with a 4XX status code were actual archived HTTP responses of missing or inaccessible pages. We also found that some URI-Ms redirected to mementos with memento-datetimes after 2016. Grusky et al. did not encounter these problems in 2016 likely due to issues relating to web archive playback, as studied by Ainsworth et al. \cite{ainsworth_only_2015} and Aturban et al. \cite{aturban_difficulties_2017}, which we worked around by discarding mementos that redirected beyond 2016.  After resolving these issues, shown in Table \ref{tab:newsroom_sample_analysis}, we were left with 277,724 mementos of news articles to evaluate.

We extracted the \texttt{META} elements from each memento with BeautifulSoup \cite{richardson_beautiful_2017}. We then extracted the field names for each field from the \emph{name}, \emph{property}, or \emph{itemprop} attribute. We assigned each field to a category by consolidating all fields in use and researched the standard or intended use case for each field. We removed instances where the field was specific to a given domain (e.g., \emph{epoch-publish-date} for \emph{bloomberg.com}) and we could find no corresponding standard. We determined the top-12 metadata categories from this process.

\section{Results and Discussion}

To address RQ1, we first examined the growth of metadata fields as a whole. Figure \ref{fig:mean-metadata-fields-over-time} demonstrates the growth in the mean number of metadata fields in use over time. In 1998, articles averaged two metadata fields. By 2006, this had risen to 17. Then the number of fields died back down to 7 in 2008 and started its consistent rising trend in 2009 with 11 fields. By 2016, the mean number of metadata fields per article was 39. This is a mean of two new fields per year between 1998 and 2016.

To further understand this trend, we categorized each metadata field by its use case (e.g., Crawler Instructions) or standard (e.g., Open Graph Protocol). We then ranked the categories of data by the number of mementos using at least one field from each category and then chose the top-12 categories. Figure \ref{fig:all_metadata_categories_streamgraph} demonstrates the growth of our top-12 metadata categories over time. Crawler instructions and Standard HTML fields were heavily used prior to 2005. In 2005, we see browser customization, specifically for Internet Explorer, added to the set of fields. By 2006, authors began to add metadata for Web Trends and Dublin Core. In 2008, authors added Search Engine Verification identifiers for Google and Bing to their pages. We see that the excitement over Internet Explorer died down by 2008, possibly due to a rise in Firefox usage and the release of Google Chrome that year \cite{w3counter_december_2008}. In the year 2009 the first article adopted Publishing Requirements for Industry Standard Metadata (PRISM). We also see the advent of search engine verification, sudden interest in Dublin Core, and continued use of Web Trends. In 2010, metadata usage exploded, supporting many different functions. Basides Standard HTML metadata, the fields with the highest adoption by 2016 were those for social cards with fields from both OGP and Twitter cards as well as those for visitor insights with Facebook Tracking. 

\begin{figure*}[htbp]
    \includegraphics[width=\textwidth]{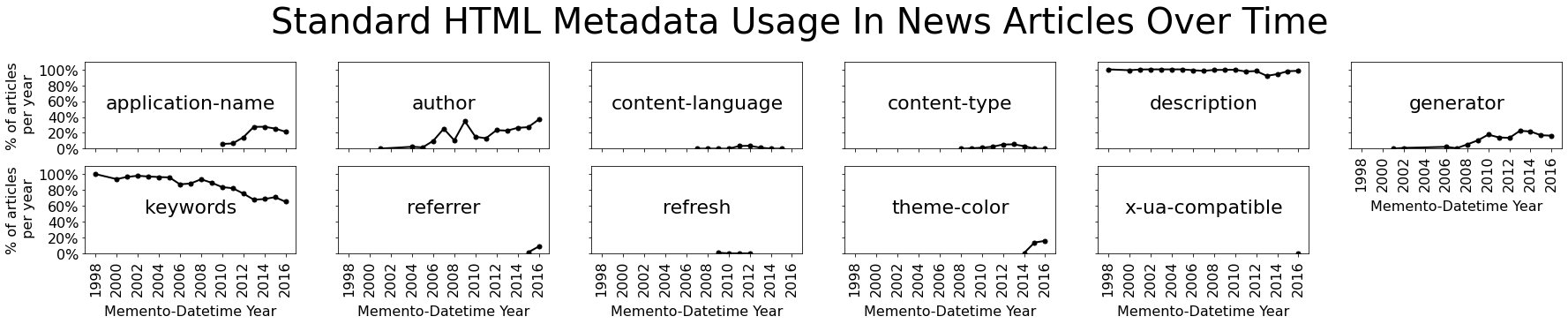}
    \caption{Use of metadata fields mentioned in HTML standards. Only fields found in the dataset are shown.}
    \label{fig:standard-html-usage}
\end{figure*}

For RQ2, we evaluated the most used metadata categories in more detail. The top five categories by number of articles are OGP, Standard HTML, Twitter Cards, Schema.org, and Facebook Tracking. Figure \ref{fig:standard-html-usage} demonstrates growing usage of several Standard HTML fields. The near-straight line across the top for \emph{description} is a feature of this dataset. All items in the dataset contain this field or other comparable description fields. The use of \emph{keywords} is in decline, possibly because of comments like those from Henzinger  \cite{henzinger_perspectives_2002} that search engines did not trust metadata. Use of the \emph{author} field is on the rise. We do know that some social card creation utilities, like embed.ly\footnote{\url{https://embed.ly}}, may honor this field. The \emph{generator} field is also on the rise, advertising the tool used to create the page (e.g., WordPress).

\begin{figure*}[htbp]
    \includegraphics[width=\textwidth]{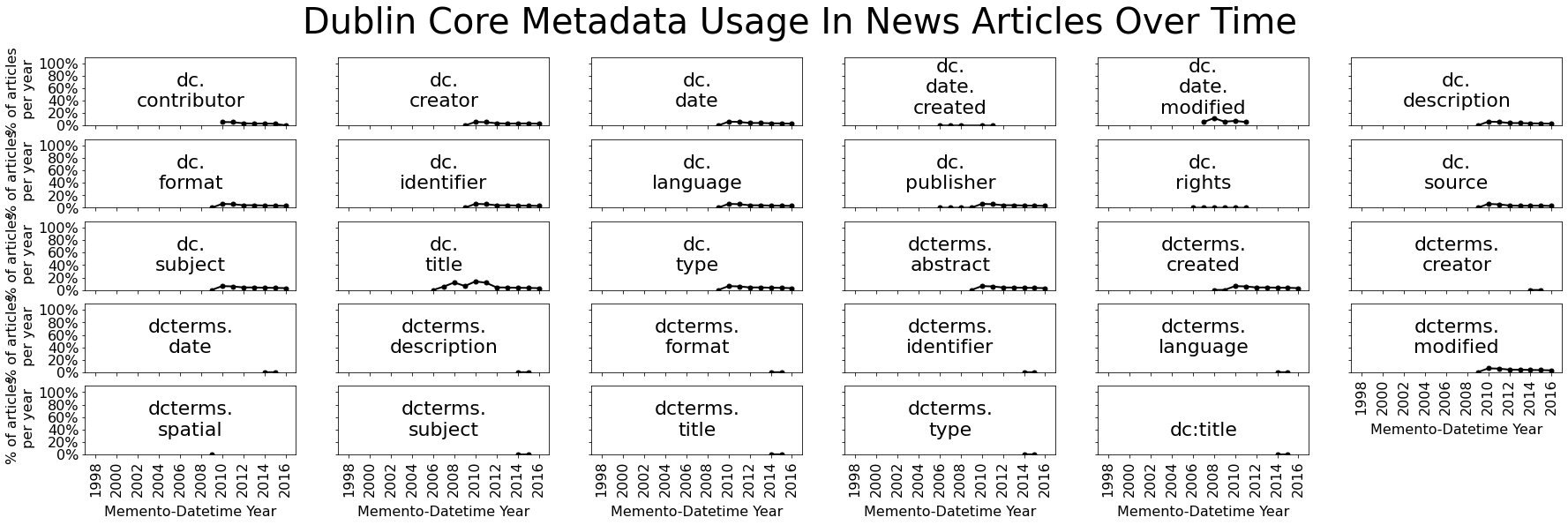}
    \caption{Dublin Core field usage in online news articles over time, showing low levels of adoption for most fields. Fields used by less than 10 articles are not shown.}
    \label{fig:dc-usage-over-time}
    \vspace{1cm}

    \includegraphics[width=\textwidth]{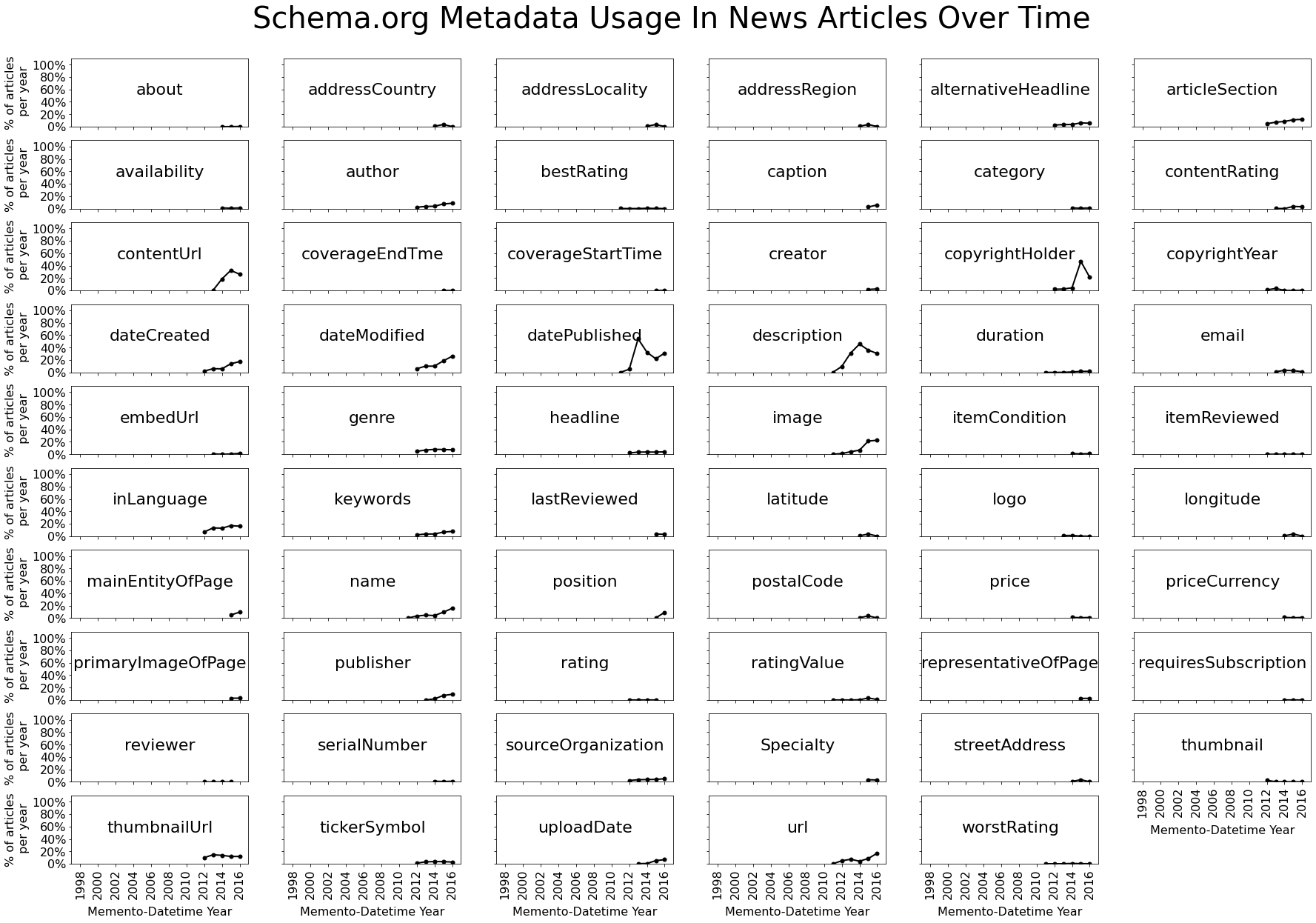}
    \caption{Schema.org field usage in online news articles over time, showing growing levels of adoption for contentUrl, description, datePublished, dateModified, and copyrightHolder, but not much else. Fields used by less than 20 articles are not shown.}
    \label{fig:schemaorg-usage-over-time}
\end{figure*}

For comparison with previous metadata studies, we also include Figure \ref{fig:dc-usage-over-time} which displays all DC fields in use by at least 10 articles. The x-axis is the memento-datetime of the article. The y-axis demonstrates the percentage of the dataset from that year that contains the field. We see very little growth of DC over time, with small peaks for certain fields like \emph{dc.title}. DC metadata was only applied by two outlets in the dataset: BBC and Fox News.

In contrast, Figure \ref{fig:schemaorg-usage-over-time} demonstrates usage of Schema.org over time in the dataset. Where DC is a general metadata standard, Schema.org has a focus on helping search engines better understand pages, potentially improving page rakings. Here we see increased usage of the \emph{contentUrl}, \emph{copyrightHolder}, \emph{dateCreated}, \emph{dateModified}, \emph{datePublished}, \emph{description}, and \emph{image} fields. The field with the highest peak is \emph{datePublished} with 54\% adoption in 2013.

\begin{figure*}[htbp]
    \includegraphics[width=\textwidth]{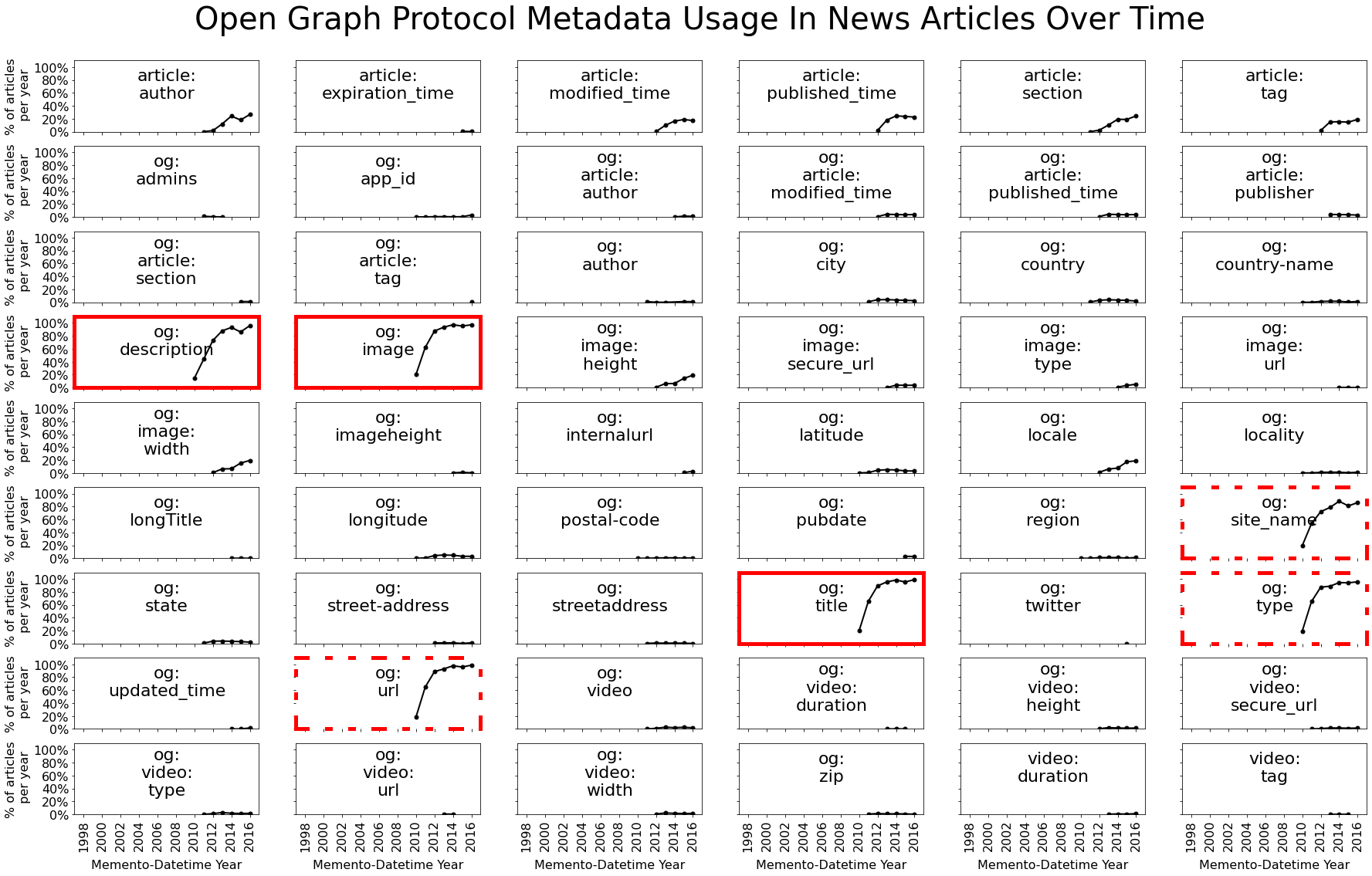}
    \caption{Open Graph Protocol field usage in online news articles over time, with fast adoption for social card fields (outlined in bold red), and fields required per the documentation (outlined in dotted red). Fields used by less than 20 articles are not shown.}
    \label{fig:ogp-usage-over-time}
    \vspace{0.1cm}

    \includegraphics[width=\textwidth]{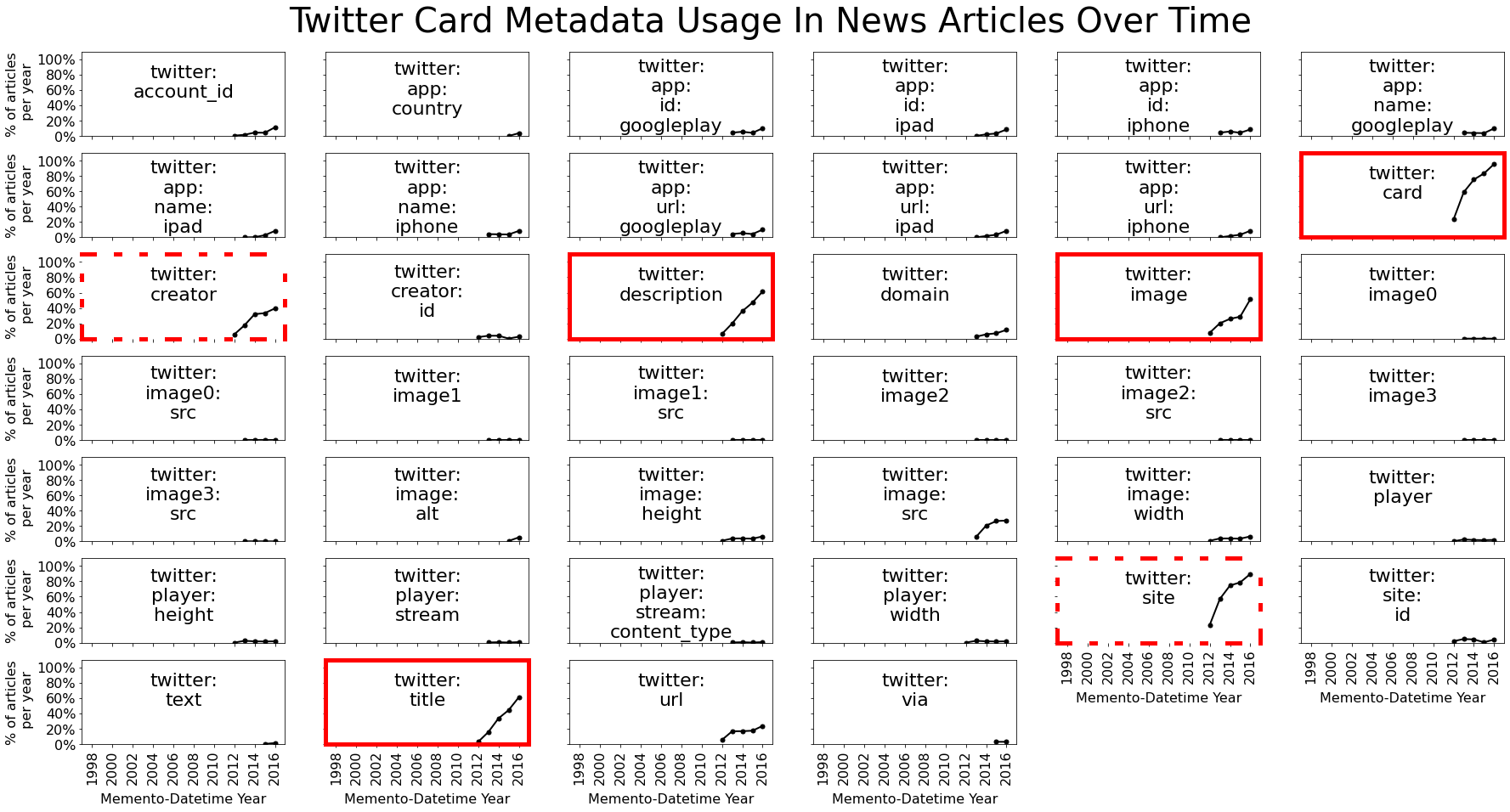}
    \caption{Twitter Card field usage in online news articles over time, with fast adoption for social card fields (outlined in bold red) and fields required per the documentation (outlined in dotted red). Fields used by less than 20 articles are not shown.}
    \label{fig:twitter-usage-over-time}
\end{figure*}

\begin{figure*}[htbp]
    \includegraphics[width=\textwidth]{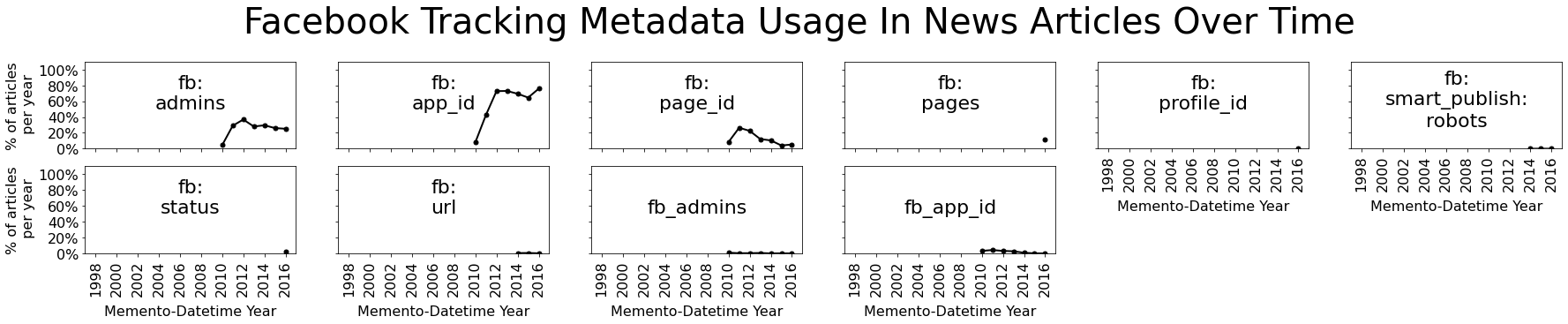}
    \caption{Facebook Tracking Metadata usage in news articles over time, showing fast adoption of \emph{fb:app\_id}.
    The fields \emph{fb\_admins} and \emph{fb\_app\_id} are included as potential archaic forms of fields not currently present in the documentation. Only fields found in the dataset are shown.}
    \label{fig:facebook_tracking_over_time}
\end{figure*}

\begin{figure}[t]
    \includegraphics[width=0.45\textwidth]{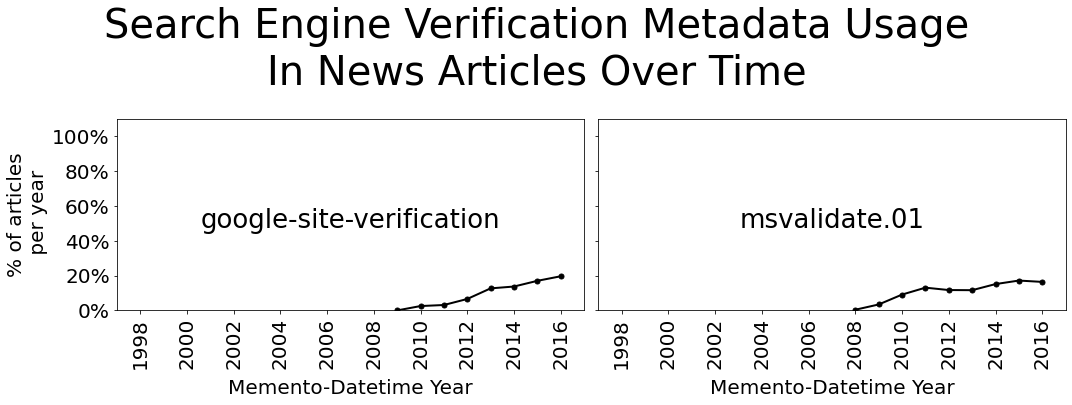}
    \caption{The use of different search engine verification fields over time.}
    \label{fig:search-engine-verification}
\end{figure}

Figures \ref{fig:ogp-usage-over-time} and \ref{fig:twitter-usage-over-time} demonstrate the changes in OGP and Twitter metadata usage over time. Where the Twitter Card standard focuses solely on how to best represent the resource within Twitter, the OGP standard also provides metadata that helps Facebook better understand the resource being shared. In spite of OGP being a more general metadata standard, news articles quickly adopted the \emph{og:title}, \emph{og:image}, and \emph{og:description} fields, all used to generate cards on Facebook. We also see a rise in usage for the \emph{og:type}, \emph{og:url}, and \emph{og:site\_name} fields. These are fields implied as required by the OGP standard, even though Facebook will still create cards without them. OGP reaches almost 20\% adoption in its first year. These fields near 80\% adoption two years later and reach 95\% adoption in 2016. A similar rise in usage occurs with \emph{twitter:card}, \emph{twitter:description}, \emph{twitter:title}, and \emph{twitter:image}. 

To further emphasize how cards appear to be driving adoption, we feature the different Facebook Tracking fields over time in Figure \ref{fig:facebook_tracking_over_time}. The top three fields are \emph{fb:app\_id}, \emph{fb:admins}, and \emph{fb:page\_id}. The \emph{fb:page\_id} field may have once been used by Facebook's Marketing API, but we could find no current documentation on its use. The closest item in current documentation is \emph{fb\_page\_id} used as an XML element in API transactions. The \emph{fb:admins} field is used to facilitate ownership of embedded comments \cite{fbcomments} on third-party pages. The \emph{fb:app\_id} field tracks page visitors \cite{fbreferral} for analytics. The \emph{fb:app\_id} field alone reaches 75\% adoption by 2016. This \emph{fg:app\_id} field is not mentioned in the OGP documentation, but Facebook's \emph{Sharing Debugger}\footnote{\url{https://developers.facebook.com/tools/debug/}} issues a warning if this field is not present. Users attempting to test their social card metadata before sharing articles on Facebook would see this warning and may add this field to their page to silence it. This notion is supported by the behavior we observed in some articles. We noticed that 165 articles provided a blank string as a value for this field and 229 articles supplied a value of \emph{FACEBOOK\_APP\_ID}. Thus, they are being encouraged to include tracking data for Facebook, but their slow adoption of other Facebook Tracking fields may indicate that it is the social cards that is driving the adoption of a field from this category, and not intentional tracking for visitor analytics. 

The adoption rates for card fields are much higher than other metadata fields. The low adoption of Schema.org in comparison to OGP and Twitter cards may be a side effect of only analyzing news articles. E-commerce sites may have much faster Schema.org adoption rates. Search engines often ask authors to include identifiers in their pages for analytics purposes. In Figure \ref{fig:search-engine-verification}, we display the growth of search engine verification fields over time. The field with the highest adoption here is \emph{google-site-verification} with 19\% of pages in 2016. Comparing this to the social card fields in Figures \ref{fig:ogp-usage-over-time} and \ref{fig:twitter-usage-over-time} indicates that sharing on social media appears to be driving metadata adoption faster than other use cases, even moreso than search engine optimization or analytics.

\section{Conclusion}

We evaluated the use of metadata in HTML news articles over time. We recognize that creating content and its metadata takes time and web page authors essentially have a metadata budget in terms of both time and effort. We sought to understand how news article authors and editors spent this budget. 

We analyzed 227,724 news articles from 29 outlets captured by the Internet Archive between 1998 and 2016. We found that in 2010, metadata usage among online news articles exploded. We see the greatest usage of metadata in the categories of Open Graph Protocol, Standard HTML, Twitter Cards, Schema.org, and Facebook Tracking. Where the mean number of metadata fields per article was two in 1998, it had grown to 37 by 2016. The rise in metadata fields is largely due to the introduction of four out of five of these categories, with only Standard HTML metadata being present since 1998.

When we break usage down by individual fields, we discover that the real motivations for metadata adoption are not for better description, search rankings, or even social media tracking, but the creation of social cards themselves. We also provide results for Dublin Core for comparison with past studies, but show little adoption of this standard in our dataset. We see more adoption of search engine verification fields each year, but this only reaches 19\% of pages by 2016. Schema.org does slightly better, with three fields peaking around 50\% adoption. Facebook tracking does better, with one field reaching 75\% adoption by 2016. None of these reach the levels of adoption of social card fields, with \emph{twitter:card}, \emph{og:title}, \emph{og:description}, and \emph{og:image} all starting around 20\% adoption in 2010 and passing 95\% adoption by 2016. On top of it, the single Facebook Tracking field reaching 75\% is itself related to social cards.

From the related work, we see a progression of studies indicating that the properties of \emph{title} and \emph{description} were already being heavily used by systems leveraging metadata. We could suppose that the growth in fields of different metadata standards corresponding to these properties is just the continuation of the same trend. But this does not explain the fast growth in fields like \emph{twitter:card}, \emph{og:image}, and \emph{twitter:image} whose sole purpose is card production.

Those, like Marchiori, who believe that metadata adoption is key to helping tools process the web have been trying to encourage web page authors to adopt different metadata standards for years. From carefully crafted expert standards like Dublin Core to standards providing perceived improvements in search rankings like Schema.org, many standards have struggled to find full adoption across news article publishers. Even social media tracking fields have not been adopted as much as they could be. Better search rankings and social media tracking is widely perceived to translate into improved revenue and better return on investment. With all of these metadata standards and the different functions they serve to choose from, what do news article authors find to be the best use of their metadata budget? It's all about the cards.

\bibliographystyle{IEEEtran}
\bibliography{references}

\end{document}